\documentclass[aps,prb,floatfix,twocolumn,showpacs,preprintnumbers,superscriptaddress,amsmath,amssymb]{revtex4}

\usepackage{graphicx}  % Include figure files
\usepackage{dcolumn}   % Align table columns on decimal point
\usepackage{bm}        % bold math
\usepackage{stmaryrd}  % for mapsfrom symbol
\usepackage{multirow}  % multirow tabular
\usepackage{color}     % color letters
\usepackage{amsmath,amssymb}
\usepackage{enumerate}
\usepackage{float}
\usepackage{ulem}

\begin{document}

%\preprint{APS/123-QED}

%\title{Effects of Coulomb and spin-orbit interactions on magnetic, topological and thermoelectric properties of zigzag nanoribbons of two-dimensional hexagonal crystals}
\title{Zigzag nanoribbons of two-dimensional hexagonal crystals:\\ magnetic, topological and thermoelectric properties}
\author{Micha\char32 l Wierzbicki} \email{wierzba@if.pw.edu.pl}
\affiliation{Faculty of Physics, Warsaw University of Technology, 00-662 Warszawa, Poland}
\author{J\'ozef Barna\'s}
\affiliation{Department of Physics, Adam Mickiewicz University, ul. Umultowska 85, 61-614 Pozna\'n, Poland}
\affiliation{Institute of Molecular Physics, Polish Academy of Sciences, ul. Smoluchowskiego 17, 60-179 Pozna\'n, Poland}
\author{Renata Swirkowicz}
\affiliation{Faculty of Physics, Warsaw University of Technology, 00-662 Warszawa, Poland}

%\date{}

\begin{abstract}
 Effects  of electron-electron and spin-orbit interactions on the ground-state magnetic configuration  and on the corresponding thermoelectric and spin thermoelectric properties in zigzag nanoribbons of two-dimensional hexagonal crystals are analyzed theoretically. Thermoelectric properties of quasi-stable magnetic states are also considered. Of particular interest is the influence of Coulomb and spin-orbit interactions on the topological edge states and on the transition between the topological insulator and conventional gap insulator states. It is shown that the interplay of both interactions has also a significant impact on  transport and thermoelectric characteristics of the nanoribbons. The spin-orbit interaction additionally determines an in-plane magnetic easy axis. Thermoelectric properties of the nanoribbons with in-plane magnetic moments are compared with those of the nanoribbons with edge magnetic moments oriented perpendicularly to their plane. Nanoribbons with ferromagnetic alignment of the edge moments are shown to reveal spin thermoelectricity, in addition to the conventional one.
\end{abstract}


\maketitle

\section{Introduction}

There is currently a great interest in strictly two-dimensional (2D) crystalline materials, like graphene,\cite{Novoselov2004}
silicene,~\cite{Aufray2010,Kara2012} germanene,~\cite{Bishnoi2013} and others. A common feature of these materials is a  2D hexagonal  lattice of atoms -- carbon, silicon, and germanium, respectively. However, both silicene and germanene have buckled atomic structure, where the two triangular sublattices are slightly displaced along the normal to the atomic plane, whereas all carbon atoms in graphene are localized in a common plane. This buckling of atomic structure leads to new properties of silicene and germanene,~\cite{Ezawa2012,Ezawa2013a,Ezawa2013b,Lado2014} which make them different from graphene.

Electronic states around the charge neutrality points of these materials are usually described by the relativistic Dirac model.\cite{Neto2009,Katsnelson2012} This model, however, neglects the electron-electron interaction. Moreover, some deviations of the exact electronic structure from the conical dispersion relation characteristic of the Dirac model appear for states distant from the Dirac points. In other words, the exact electronic structure is more complex, and either tight binding models or  {\it ab-inito} techniques are usually employed to calculate the corresponding band structure, especially in the case of nanoribbons.\cite{Lopez2013,Zhang2010,Lou2009,Lou2011,Guan2012,Lou2013,Chen2014,Ding2013}
It is already well known that the Coulomb interaction between electrons leads to edge magnetism in nanoribbons with zigzag edges.\cite{Yazyev2010} In the ground state of the system, magnetic moments localized at one edge are opposite to those at the other edge.\cite{Yazyev2010,Son2006,Pisani2007}  In some situations, however, one can stabilize a quasi-stable ferromagnetic state, where all edge moments are oriented in parallel.\cite{Haugen2008,Kim2008,Zheng2012,Li2014,Wang2014} Spin-orbit interaction gives rise to a magnetic anisotropy, and when this coupling is  sufficiently strong,
an in-plane orientation of the edge moments is energetically more favorable.\cite{Lado2014}
Thus, the interplay of spin-orbit and Coulomb interactions is of fundamental importance for the edge magnetism.

The spin-orbit interaction also leads to topologically protected edge states in the spin-orbit gap,\cite{Kane2005}
and these states
are responsible for topological-insulator properties of these materials.\cite{Hasan2010,Qi2011} However, Coulomb interaction has a significant impact on these properties.\cite{} Accordingly, the interplay of spin-orbit and Coulomb interactions is important not only for the edge magnetism, but also  for the topological state of these materials.
Therefore, in this paper we consider this issue in the case of zigzag nanoribbons of two-dimensional buckled hexagonal crystals, and especially the role of Coulomb interaction and spin-orbit coupling.
We focus on the thermoelectric and spin thermoelectric effects of these materials in the ballistic transport regime, and on their dependence on the Coulomb and spin-orbit interactions.  First, we calculate energy of various magnetic configurations in  zigzag nanoribbons, and find that the antiparallel configuration with in-plane magnetic moments is the ground state structure for large enough spin-orbit coupling. For small spin-orbit coupling, the  in-plane and perpendicular-to-plane configurations are degenerate. Then, we calculate the electronic states, with a special emphasis  on the edge states responsible for the topological properties. Having determined the magnetic state and also the electronic edge states, we calculate thermoelectric properties with the main focus on the role of topological edge states and transition from the topological to conventional insulator state.

Generally, the thermoelectric and especially the spin thermoelectric effects in systems of reduced dimensionality are currently of great interest,\cite{Hochbaum2008,Harman2002,Duarte2009,Walter2011,Liebing2011,Uchida2008,Swirkowicz2009,Misiorny2014} mainly due to a hope to find an efficient way to  convert  dissipated heat into electrical energy.
In fact, thermoelectric properties of nanoribbons have been studied theoretically in a couple of papers.\cite{Pan2012,Zberecki2013,Yan2013,Zberecki2014,Zberecki2014a} However, the role of Coulomb interaction and topological edge states in the gap, has not been studied thoroughly enough yet. In a recent paper\cite{Wierzbicki2015} we have analyzed the influence of topological states on the thermoelectric properties, with the main focus on the role of a staggered exchange field~\cite{Ezawa2012,Ezawa2013a,Ezawa2013b} and of an electric field perpendicular to the atomic plane. Here we consider in more details the role of Coulomb interaction, which has a significant impact on  the topological properties.\cite{Matthes2014}

In section 2 we present the Hamiltonian used to describe the 2D materials under consideration, and also introduce briefly the method used to calculate transmission function and then all the thermoelectric and transport coefficients. In section 3 we present numerical results on the edge magnetism, topologically protected states, and thermoelectric properties of the corresponding zigzag nanoribbons. Summary and final conclusions are in section 4.

\section{Theoretical description}

\subsection{Tight-binding Hamiltonian}

The zigzag nanoribbons of hexagonal 2D crystals, like silicene or germanene, which are analyzed in this paper consist of $N$ zigzag atomic chains.
The system can be described by Hamiltonian of the following general form:
\begin{equation}
H=H_{\rm tb}+H_{\rm so}+H_{\rm e-e},
\label{eq01}
\end{equation}
where the first term, $H_{\rm tb}$, is the tight-binding Hamiltonian, the term $H_{\rm so}$ represents the spin-orbit
coupling, while the last term, $H_{\rm e-e}$, stands for the electron-electron Coulomb interaction.

The tight-binding Hamiltonian is assumed in the form
\begin{equation}
H_{\rm tb}=-t\,\sum\limits_{\langle i,j\rangle}\,c_{i\sigma}^\dagger\,c_{j\sigma},
\label{eq02}
\end{equation}
where $c^\dagger_{i\sigma}$ ($c_{j\sigma}$) is the creation (annihilation) operator for
an electron with spin $\sigma=\uparrow,\downarrow$ at the lattice point $i$ ($j$). The spin quantization axis is taken along the $z$-axis, which is normal to the atomic plane.
The tight-binding term is restricted here to electron hopping between nearest-neighbor sites, with the corresponding
hopping parameter $t$. For instance, $t=1.6\ \mbox{eV}$ in silicene.\cite{Ezawa2012}

The intrinsic spin-orbit interaction
in Eq.~(\ref{eq01}) follows from electron hopping between next-nearest-neighbors in the hexagonal lattice,
and can be written in the form,\cite{Ezawa2013b}
\begin{equation}
H_{\rm so}=i\lambda_{\rm so}\,
\sum\limits_{\langle\langle i,j\rangle\rangle}
\nu_{ij}\,
\left(
c_{i\uparrow}^\dagger\,\,
c_{j\uparrow}-
c_{i\downarrow}^\dagger\,\,
c_{j\downarrow}
\right),
\label{eq02so}
\end{equation}
where $\lambda_{\rm so}$ is the spin-orbit interaction parameter, while
$\nu_{ij}= 1$  ($\nu_{ij}= -1$) when the hopping path from
second-neighbor sites $i$ to $j$ in the hexagonal lattice is clockwise
(anticlockwise) with respect to the positive $z$-axis.

The electron-electron interaction will be
taken into account in the Hubbard form within the mean-field approximation (MFA). The  form of the MFA Hubbard Hamiltonian depends on whether the spin quantization axis is parallel to the local magnetization orientation or not. We assume, similarly as in Ref.[\onlinecite{Lado2014}], that in a hypothetical state the magnetic moments are tilted from the normal orientation by an angle $\alpha$, and will calculate the corresponding total energy. The ground state configuration will be thus the one with the lowest energy. Assuming the quantization axis parallel to the local magnetic moments, the Hubbard term in the mean-field approximation can be written in the form
\begin{equation}
H_{\rm e-e}=U\sum\limits_i \left(
n^{\alpha }_{i\uparrow}\left<n^{\alpha }_{i\downarrow}\right>
+n^{\alpha }_{i\downarrow}\left<n^{\alpha }_{i\uparrow}\right>
-\left<n^{\alpha }_{i\uparrow}\right>\left<n^{\alpha }_{i\downarrow}\right>
\right),
\label{eqint}
\end{equation}
where $n^{\alpha }_{i\uparrow}$ and $n^{\alpha }_{i\downarrow}$ are the particle number operators for a site $i$ and for electrons with spin $\uparrow$ and $\downarrow$ with respect to the spin quantization axis tilted from the normal to the plane towards a nanoribbon edge by an angle $\alpha$. In turn,
$U$ is the on-site Coulomb repulsion parameter, $U>0$. We also note, that the Coulomb interaction of electrons at different lattice sites is omitted.

The MFA Hubbard Hamiltonian should be now written in the quantization axis common for all terms of the Hamiltonian. This can be done by rotating the quantization axis back to the orientation normal to the plane, which is achieved by the transformation\cite{Swirkowicz1997}%
\begin{equation}
c^\alpha_{i\sigma}=\cos\left(\frac{\alpha}{2}\right)\,c_{i\sigma}-\tilde\sigma\,
\sin\left(\frac{\alpha}{2}\right)\,c_{i,-\sigma}
\end{equation}
\noindent
where  $\tilde\sigma=1$ ($\tilde\sigma=-1$) for $\sigma=\uparrow$ ($\sigma=\downarrow$).

The spin-dependent mean number of electrons, $\left<n^\alpha_{i\sigma}\right>$, at the site $i$,
required to evaluate the Hubbard interaction term (\ref{eqint}),
is determined self-consistently from the band-structure of the corresponding nanoribbon,
\begin{equation}
\left<n^\alpha_{i\sigma}\right>=\sum\limits_{\rm bands}\frac{a}{2\pi}\int\limits_0^{2\pi/a} n^\alpha_{i\sigma}(k)\,
f[E(k),\mu]
\, dk ,
\label{eq03}
\end{equation}
where $f(E,\mu)$ is the Fermi-Dirac distribution function, with $\mu$ denoting the chemical potential, while $n^\alpha_{i\sigma}(k)$
is the spin-dependent electron density at site $i$ for a Bloch-wave of energy $E(k)$.
The parameter $a$ denotes the length of the elementary cell of a zigzag nanoribbon.
The chemical potential of a pristine nanoribbon is equal to the corresponding Fermi energy $E_F$.
By doping with donor or acceptor impurities, or by applying an external gate voltage, one can shift up or down the Fermi level, and its  position with respect to $E_F$ will be described by $\mu-E_F$. In the following $\uparrow$ and $\downarrow$ will denote spin projection on the quantization axis parallel to the orientation of the magnetic moments.

To identify the edge-states, we evaluate the
mean value of $\xi=2(y-y_0)/w$,
where the $y$-axis is in the nanoribbon plane and normal to the nanoribbon axis,
$y_0$ is the $y$-coordinate of the nanoribbon center, while $w$ is the width of the nanoribbon.
For $\xi=-1$ ($\xi=1$), the states are localized at the
left (right) edge of the nanoribbon.

\subsection{Electronic transmission and general formulas for thermoelectric coefficients}

Ballistic electronic transmission in quasi-one-dimensional structures is usually determined in terms of the Green's functions
formalism and an expression derived by Caroli {\it et al}.\cite{Caroli71} An alternative
approach is based on matching wave functions in the scattering region to the Bloch modes of ideal bulk leads.\cite{Khomyakov2005}
The total transmission, $T_{\rm tot}(E)$, as a function of electron energy $E$ is then defined
as the sum of transmissions for all incident Bloch waves for a given wave-number $k$,
and is a piece-wise constant function
equal to the number of energy bands for a given
energy $E$.
In turn, the spin transmission,
$T_{\rm spin}$, is equal to the sum of the mean values of the Pauli operator $\hat\sigma_z$.
More detailed description of the transmission calculation is presented in Ref. [\onlinecite{Wierzbicki2015}].
The total transmission, $T_{\rm tot}$,  and spin transmission, $T_{\rm spin}$, are related to the spin-dependent transmission, $T_\sigma$ ($\sigma=\uparrow,\downarrow$), as
$T_{\rm tot}=T_{\uparrow}+T_{\downarrow}$ and
$T_{\rm spin}=T_{\uparrow}-T_{\downarrow}$, respectively.

Knowledge of the transmission function $T_\sigma(E)$  allows to calculate all transport characteristics
in the linear response regime, including also the thermoelectric properties.
The spin resolved transport and thermoelectric coefficients can be then determined
from the moments $L_{m\sigma}$ of the electronic current series expansion around the Fermi level,
\begin{equation}
L_{m\sigma}=\frac{1}{h}\int\limits_{-\infty}^{\infty}
T_\sigma(E)\,\left(E-\mu\right)^m\,\frac{\partial f(E,\mu)}{\partial E}\,dE.
\label{eq04}
\end{equation}
For a piece-wise
constant transmission function, calculation of $L_{m\sigma}$ can be reduced to
calculation of the
so-called
incomplete Fermi-Dirac integral,\cite{Wierzbicki2015}
\begin{equation}
{\cal F}_m(x)=\frac{1}{m!}\,\int\limits_x^\infty \frac{z^m\,dz}{1+e^z}\ ,\quad\mbox{for}\ x\ge 0,
\end{equation}
for which there exists an efficient numerical procedure.\cite{goano}

The spin-dependent conductance $G_\sigma$ is proportional to
the moment $L_{0\sigma}$, $G_\sigma=e^2\,L_{0\sigma}$
where $e$ is the electron charge. The total conductance $G$ is then equal to $G_\uparrow+G_\downarrow$, and the spin conductance is $G_s=G_\uparrow-G_\downarrow$.
The heat conductance is defined at vanishing charge current and can be written as
$
\kappa=(1/T)\sum_\sigma (L_{2\sigma}-L_{1\sigma}^{2}/L_{0\sigma})
$,
where $T$ stands for temperature.
In turn, the spin-dependent thermoelectric coefficient (thermopower)
$S_\sigma$ can be expressed in the form
\begin{equation}
S_\sigma=-\frac{1}{|e|T}\frac{L_{1\sigma}}{L_{0\sigma}}
\end{equation}
for $\sigma=\uparrow,\downarrow$.
Then, one can define the conventional (charge) Seebeck coefficient (thermopower) as $S_c=\frac{1}{2}(S_\uparrow+S_\downarrow)$ and
also the corresponding spin Seebeck coefficient (spin thermopower) as $S_s=\frac{1}{2}(S_\uparrow-S_\downarrow)$.\cite{Swirkowicz2009}
When the spin thermopower is absent, one can write the conventional Seebeck coefficient $S$ ($S_c\to S$) in the standard form
$S=-(1/|e|T)(L_{1\uparrow}+L_{1\downarrow}/L_{0\uparrow}+L_{0\downarrow})=
-(1/|e|T)(L_1/L_0)$,
with $L_n=L_{n\uparrow}+L_{n\downarrow}$ for $n=0,1,2$. In turn,  the electronic contribution $\kappa$ to the thermal conductance is given by
$\kappa=(1/T)(L_2-L_1^2/L_0)$.

Finally, one can define the dimensionless
figures of merit for the conventional and spin thermoelectric effects,
\begin{subequations}
\begin{align}\label{eq-ZTc}
ZT_c=\frac{S_c^2\,G\,T}{\kappa+\kappa_{\rm ph}}, \\
\label{eq-ZTs}
ZT_s=\frac{S_s^2\,\vert G_s\vert \,T}{\kappa+\kappa_{\rm ph}},
\end{align}
\end{subequations}
where $\kappa_{\rm ph}$ is the phonon contribution to the heat conductance.
In the absence of spin thermoelectricity, only the
dimensionless figure of merit for the conventional thermoelectricity is relevant ($ZT_c\to ZT$), and
$ZT=S^2\,G\,T/(\kappa+\kappa_{\rm ph})$,

\section{Numerical results}

In this section  we present some numerical results on the basic thermoelectric properties.
Before this, however, we consider
various stable and quasi-stable magnetic states of the nanoribbons. Then, we consider separately the antiferromagnetic (AFM) states, with magnetic moments at one edge being antiparallel to the magnetic moments at the other edge), and ferromagnetic (FM) states, with parallel alignment of all the edge moments.

\subsection{Stability diagram}

We have performed numerical calculations of the total energy of nanoribbons with the edge magnetic moments tilted by an angle $\alpha$
from the orientation normal to the nanoribbon plane, for $\alpha$ ranging from $\alpha = 0$  to $\alpha = \pi /2$. The calculations have been carried out for both FM and AFM magnetic configurations.
The numerical results clearly show that the states for $\alpha\ne 0,\pi /2$ are neither stable nor quasi-stable. The only stable or quasi-stable states are those for $\alpha = 0$ (moments normal to the plane) and $\alpha = \pi /2$ (moments in the nanoribbon plane). This applies to both FM and AFM  configurations. From this analysis, we have constructed the stability diagram of the FM and AFM states.
Due to the spin-orbit interaction, both AFM and FM configurations have lower energy when the edge moments are oriented in the nanoribbon plane than when they are along the  normal to the plane, and the energy difference depends on the Hubbard parameter $U$ and spin-orbit parameter $\lambda_{\rm so}$.
Thus, for each magnetic configuration one can distinguish between the perpendicular ($\perp$) and parallel ($\parallel$) states.\cite{Lado2014} In the former case the edge moments are normal to the nanoribbon plane, while in the latter one they are in the nanoribbon plane.
As a result, one finds the following four states, FM$\perp$, AFM$\perp$, FM$\parallel$ and AFM$\parallel$.
Numerical results show that the AFM$\parallel$ state is the ground state of the system, i.e. the state of the lowest energy for $\lambda_{\rm so}>0$.

\begin{figure}[h]
\null\hskip 0.02\columnwidth
\includegraphics[width=0.97\columnwidth]{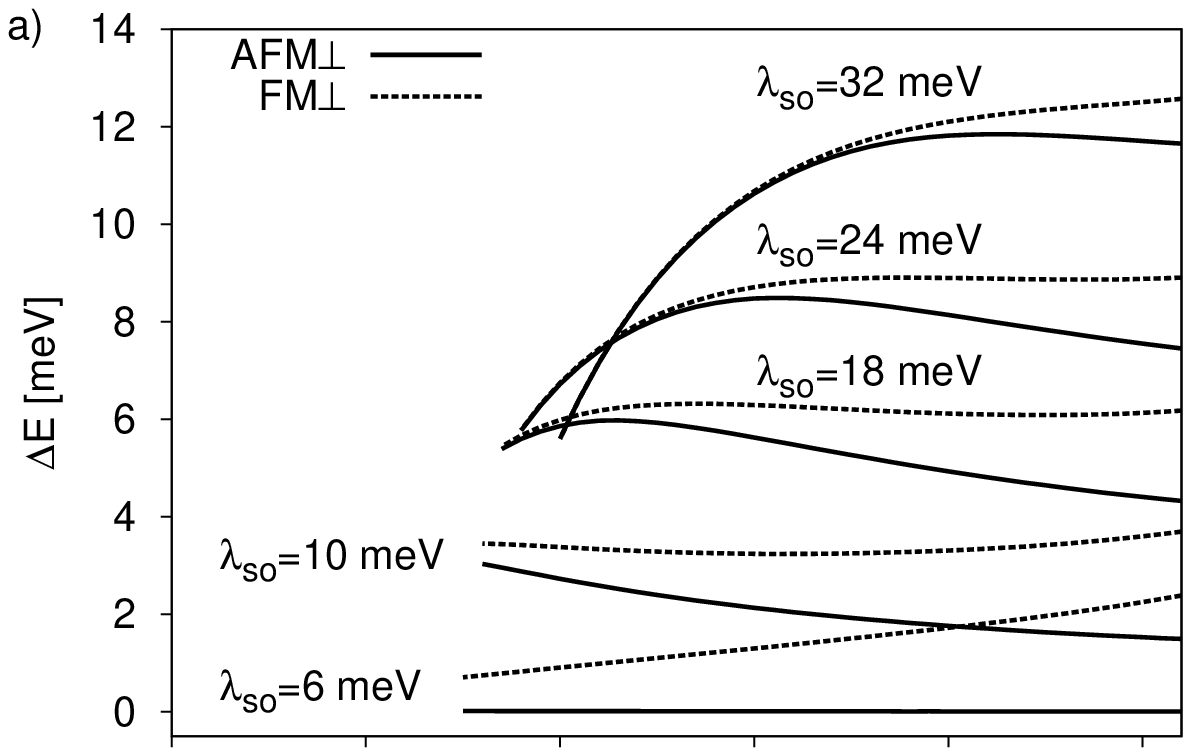}
\includegraphics[width=0.99\columnwidth]{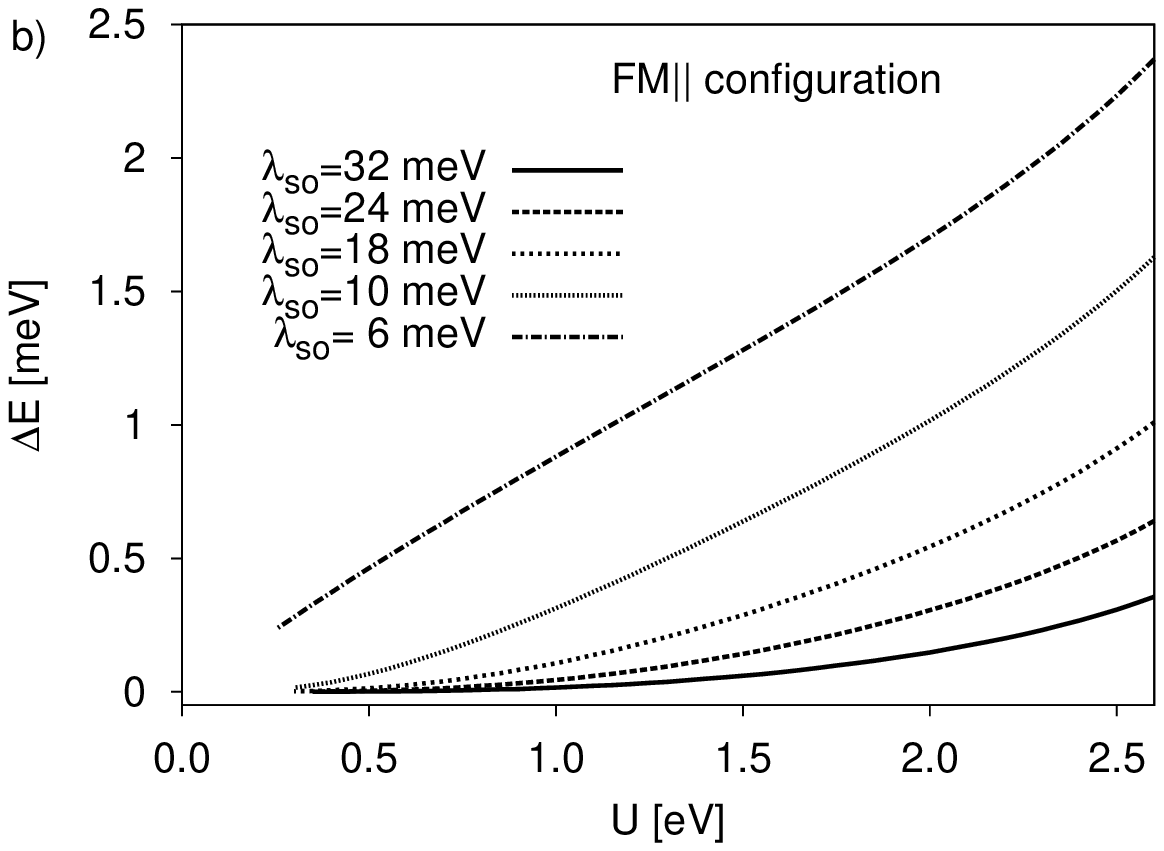}
\caption{\label{fig1}Energy difference, $\Delta E$, of the a) AFM$\perp$, FM$\perp$, and b) FM$\parallel$ states
and of the AFM$\parallel$ ground state, presented
as a function of the Hubbard parameter $U$ for indicated values of the spin-orbit parameter $\lambda_{\rm so}$. Other parameters: $t=1.6$ eV, and $N=8$.}
\end{figure}

In Fig.~\ref{fig1} we show the energy ${\Delta}E$ of the states FM$\perp$, AFM$\perp$, and  FM$\parallel$, measured from the energy of the AFM$\parallel$ ground state.  The excess energy, ${\Delta}E$, of a particular state over the
AFM$\parallel$ ground state is presented there as a function of $U$ for indicated values of the spin-orbit parameter $\lambda_{\rm so}$.
One should note, that only nonmagnetic solutions have been found for low values of $U$,
while magnetic phases appear for larger values of $U$.
>From Fig.~\ref{fig1}a follows, that the perpendicular configuration of magnetic moments
is not favorable in the presence of spin-orbit coupling,
which is in agreement with the results obtained by Lado {\it et al}.\cite{Lado2014}
Both AFM$\perp$ and FM$\perp$ states correspond to energies higher than that of the AFM$\parallel$ state,
and the energy difference strongly increases with the spin-orbit parameter.
As one might expect, the AMF$\perp$ and AMF$\parallel$ states
are practically degenerate in the limit of small values of $\lambda_{\rm so}$.
In turn, energy of the FM$\perp$ state is considerably higher than the energy of the AMF$\perp$ state, and the difference increases roughly linearly with increasing $U$.
Interestingly, the energy difference between the perpendicular phases AFM$\perp$ and FM$\perp$ for a given $U$ decreases
with increasing $\lambda_{\rm so}$, so these two configurations become almost degenerate for strong spin-orbit coupling and
moderate  values of the Hubbard parameter $U$. Remarkably different behavior can be observed for the FM$\parallel$ state, as shown in Fig.~\ref{fig1}b.
For high values of $\lambda_{\rm so}$ and moderate values of $U$, this state is practically degenerate  with the AFM$\parallel$ one,
but the energy of FM$\parallel$ state notably increases when the spin-orbit interaction becomes small.
Finally, in the limit of $\lambda_{\rm so}\rightarrow 0$, the states AFM$\parallel$ and AFM$\perp$ are fully degenerate
for all relevant values of $U$ (Fig.~\ref{fig1}a), whereas the FM states are quasi-stable and correspond to higher values of energy.

Below we will analyze transport and thermoelectric properties of the AFM$\parallel$ ground state as well as of the quasi-stable states  AFM$\perp$, FM$\perp$, and  FM$\parallel$. Though the latter states correspond to energy larger than the ground-state one, they can be stabilized  by some external forces, like a weak magnetic field, exchange coupling to an adjacent  magnetic subsystems, etc.

\subsection{Phonon heat conductance}

To calculate some of the thermoelectric parameters, like thermoelectric efficiency (figure of merit),
one needs to know not only the electronic contribution $\kappa$ to the heat conductance, but also the corresponding phonon term $\kappa_{\rm ph}$.
In a recent paper Yang {\it et al}~\cite{Yang2014} have determined the phonon contribution to the heat conductance
of narrow germanene and silicene nanoribbons,
utilizing the Landauer formula (ballistic transport approximation) and the
phonon transmission functions calculated by an ab-initio method.
We checked that $\kappa_{\rm ph}$ presented in Ref.[\onlinecite{Yang2014}]
scales linearly with the nanoribbons width, and
then extrapolated $\kappa_{\rm ph}$ up to zigzag nanoribbons containing $N=8$ zigzag chains
(16 atoms in the zigzag chain), which are considered in this paper.
We obtained the following values for $N=8$  and $T=100\ \mbox{K}$:
$\kappa_{\rm ph}=0.92\ \mbox{nW/K}$ for germanene, and $\kappa_{\rm ph}=1.09\ \mbox{nW/K}$
for silicene. Thus, for the model calculations we take $\kappa_{\rm ph}\approx 1\ \mbox{nW/K}$
in Eqs (\ref{eq-ZTc}) and (\ref{eq-ZTs}).

\subsection{Antiferromagnetic phase}

\begin{figure}[h]
\includegraphics[width=0.55\columnwidth]{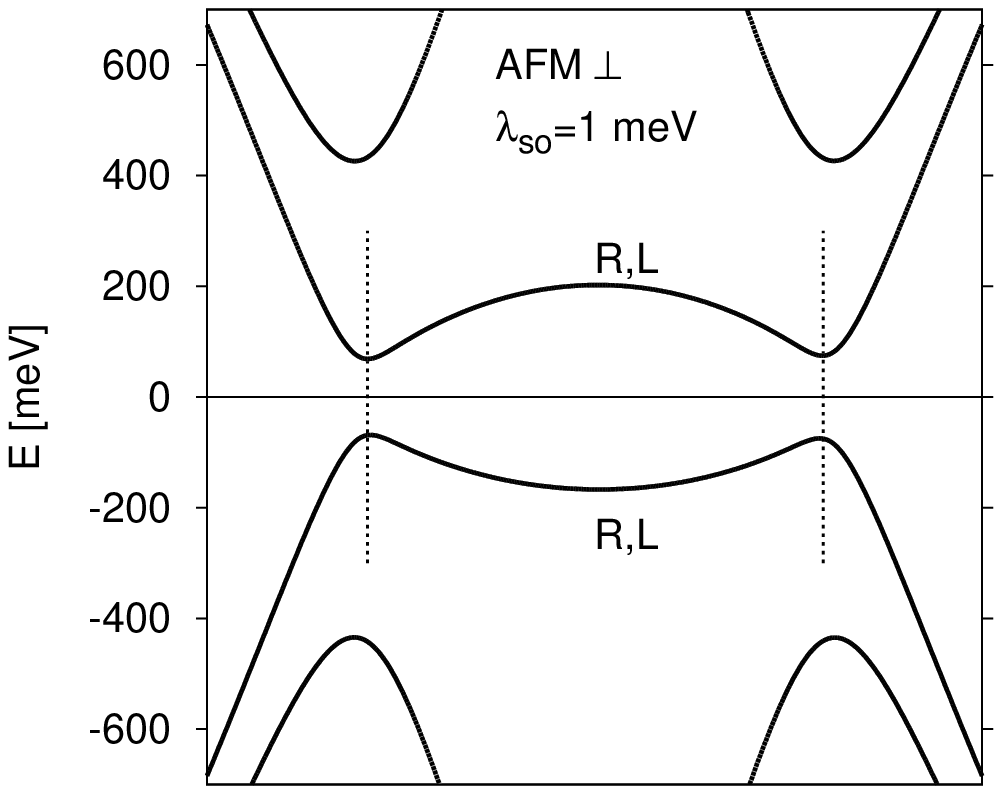}
\includegraphics[width=0.43\columnwidth]{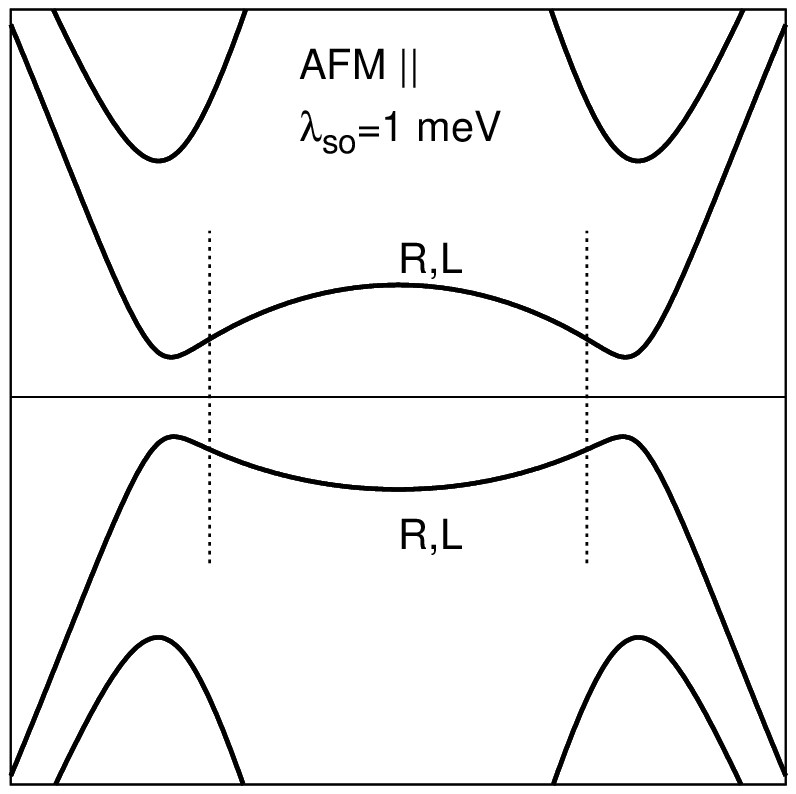}
\includegraphics[width=0.55\columnwidth]{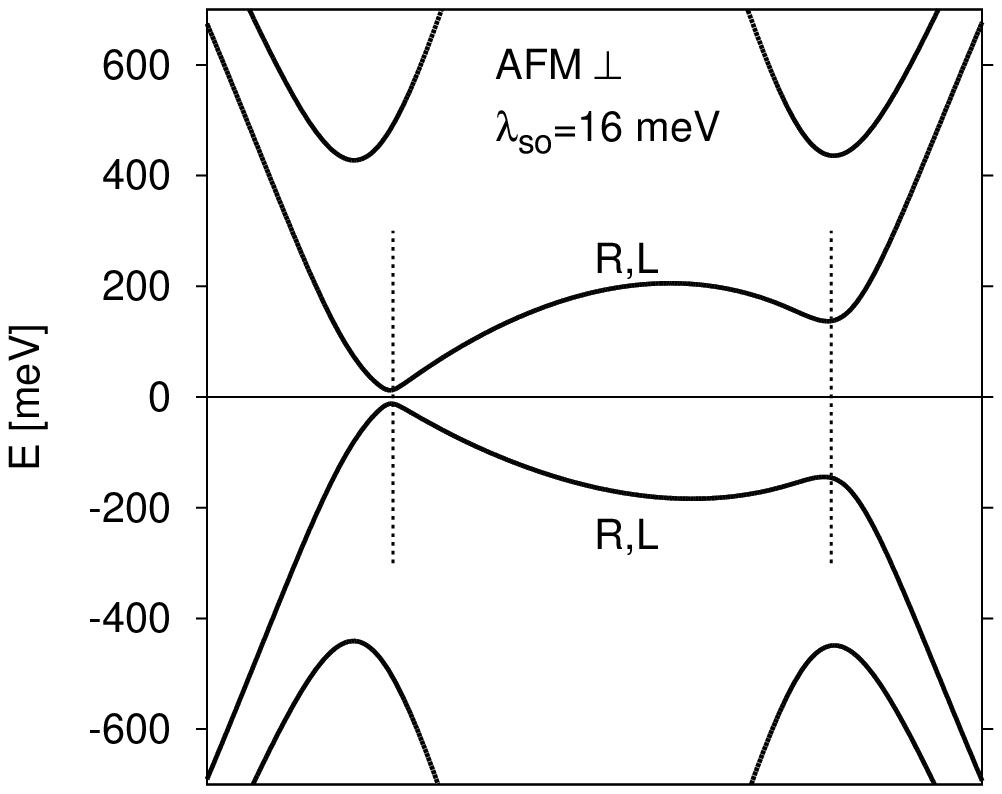}
\includegraphics[width=0.43\columnwidth]{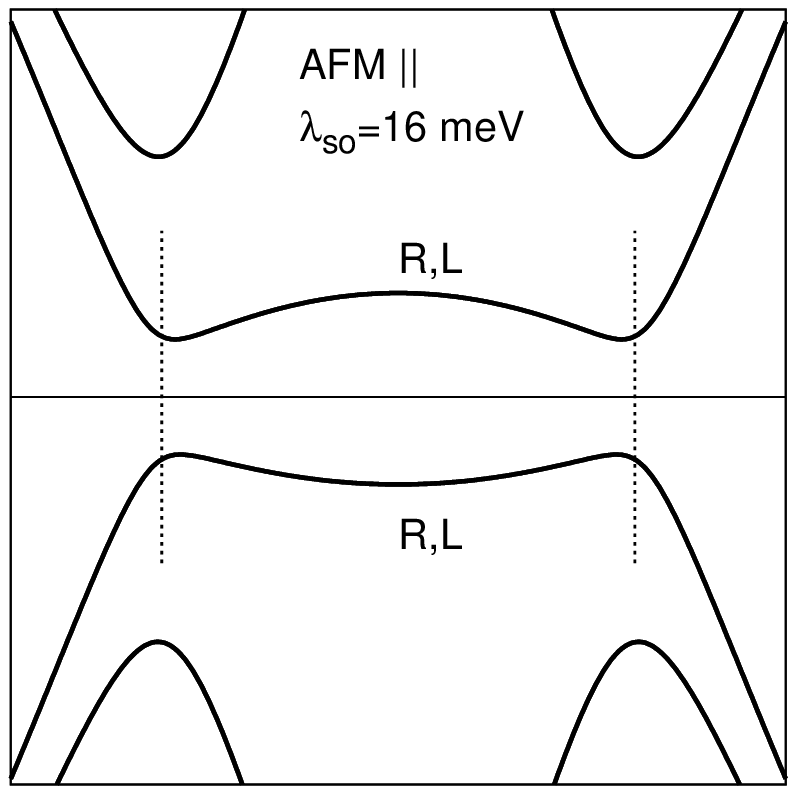}
\includegraphics[width=0.55\columnwidth]{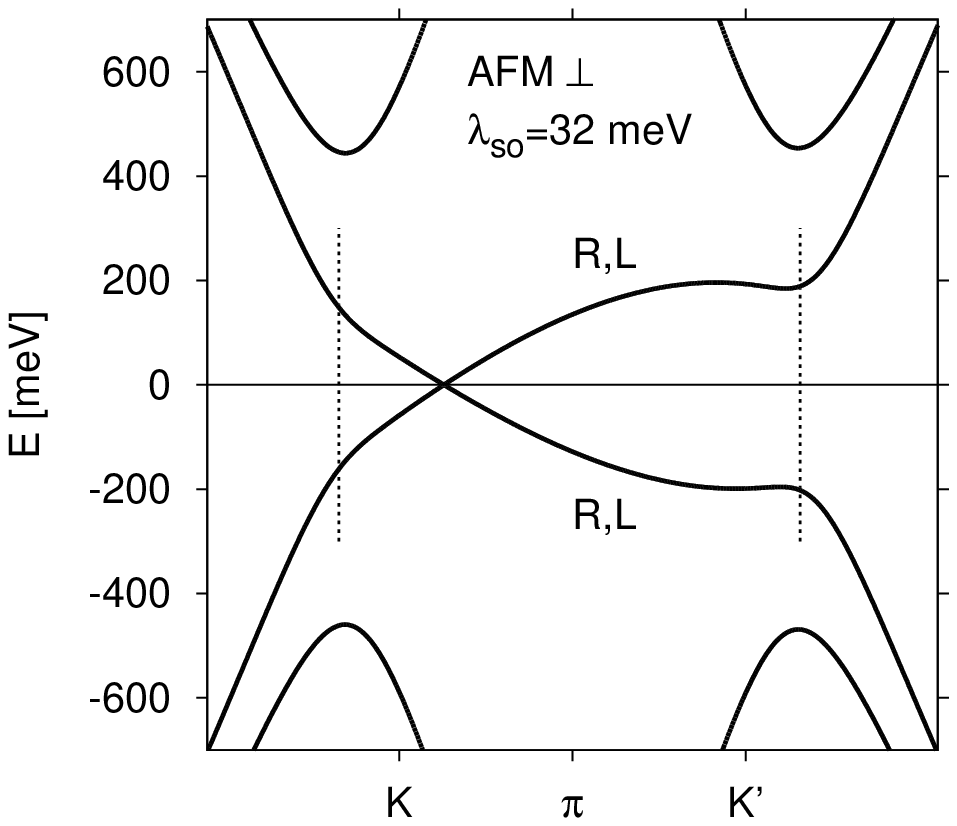}
\includegraphics[width=0.425\columnwidth]{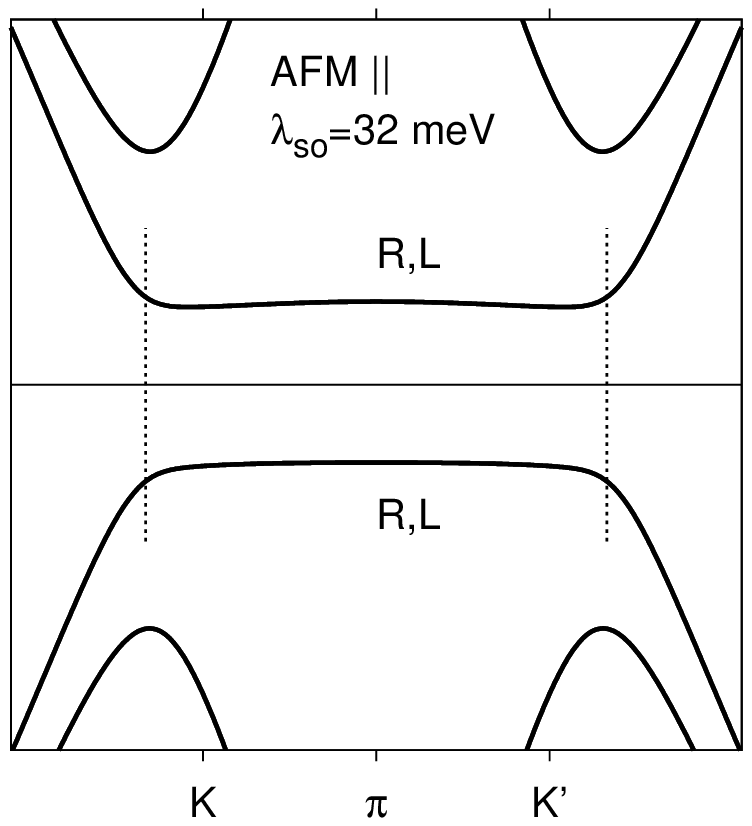}
\caption{\label{fig2} Electronic structure of the AMF$\perp$ (left panel) and AMF$\parallel$ (right panel) states for indicated values of the spin-orbit parameter.
Vertical dotted lines indicate the interval in the wavevector space, where the edge parameter $\xi$ obeys the condition, $|\xi|>0.5$. The other parameters are: $U=1.4$~eV, $N=8$, and $t=1.6$~eV.}
\end{figure}

\begin{figure}[h]
\includegraphics[width=0.99\columnwidth]{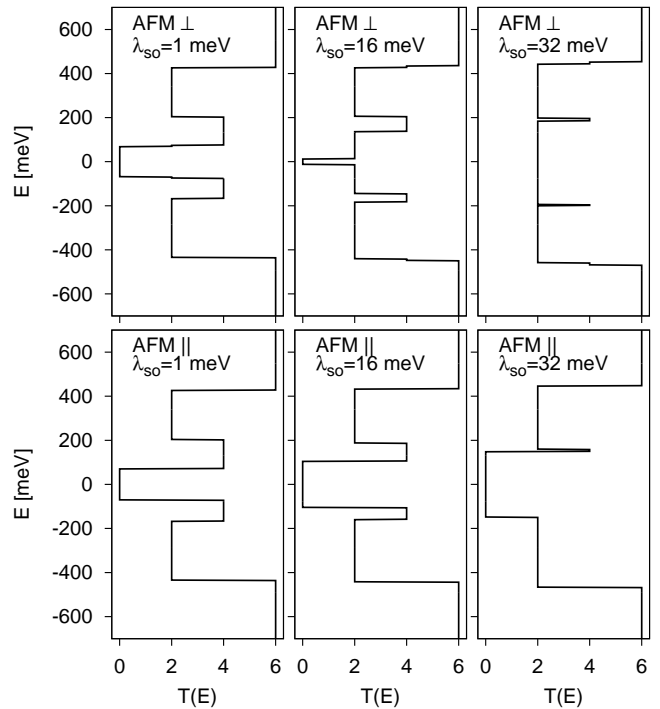}
\caption{\label{fig3} Transmission function for the AFM$\perp$ (top panel) and AFM$\parallel$ (bottom panel) phases, calculated for the same parameters as in Fig.~\ref{fig2}.}
\end{figure}

Now, we analyze ballistic transport and thermoelectric properties in the AFM configuration in more details. The corresponding electronic states
for the AMF$\perp$ (left panel) and AMF$\parallel$ (right panel) configurations are presented in Fig.~\ref{fig2}
for three different values of the spin-orbit parameter.
The corresponding transmission functions  are presented in Fig.~\ref{fig3}. The lowest value of $\lambda_{\rm so}$ assumed in these figures
is comparable to the spin-orbit parameter typical for silicene\cite{Ezawa2012b}
(note the factor $3\sqrt{3}$ in the definition of spin-orbit coupling  in this reference).
However, larger values of the parameter $\lambda_{\rm so}$ can appear in other related two-dimensional crystals, like germanene\cite{Liu2011}
($\lambda_{\rm so}=43\ \mbox{meV}$) or stanene\cite{Liu2011} ($\lambda_{\rm so}=29.9\ \mbox{meV}$).
Moreover, results for larger values of the spin-orbit parameter are very  interesting from the point of view of the topological properties.
Therefore, in numerical calculations of the band structure, transmission, and thermoelectric coefficients
we assume the following three values of $\lambda_{\rm so}$: $\lambda_{\rm so}=0.02 t=32$~meV (strong spin-orbit coupling),
$\lambda_{\rm so}=0.01 t=16$~meV, and $\lambda_{\rm so}=1$~meV (weak spin-orbit coupling).
In turn, for the electron-electron interaction we assume $U=1.4$~eV.

To emphasize behavior of the edge states, which are responsible for topological properties, we displayed in Figs.~\ref{fig2} and \ref{fig3}
only a small energy region near the Fermi level $E_F$ of the corresponding charge-neutral nanoribbon.
Moreover, only the region of wavevectors $k$, in which the states are strongly localized near the left (L) or right (R ) edges
of the nanoribbons is presented there. As follows from these figures, the electronic spectra reveal an insulating gap for small values of $\lambda_{\rm so}$, and this  gap appears
in both AFM$\perp$ and AFM$\parallel$ phases. Moreover, width of the  gap
is practically the same in both cases. Note, such a gap is absent in the limit of $U=0$, when zero-energy topologically protected states  appear.\cite{Wierzbicki2015}
Furthermore, the valence and conduction states localized at the left and right edges are degenerate  in both phases.
These states are additionally spin degenerate.

Qualitatively different behavior can be observed for larger values of $\lambda_{\rm so}$.
The edge states are then still degenerate, but in the AFM$\perp$ phase they cross the Fermi level,
so the system acquires properties of a topological insulator,  whereas the AFM$\parallel$ phase still exhibits properties of a conventional insulator.
Thus, at a certain  value of $\lambda_{\rm so}$, there is a transition from the conventional to topological insulator in the AFM$\perp$  phase, while no such a transition appears in the AFM$\parallel$ phase. Moreover, one may conceive that by applying an appropriate external magnetic field
it would be possible to switch between the AFM$\parallel$ and AFM$\perp$ states, and thus also to change the character
of the system from conventional to topological insulator, and {\it vice versa}.
Opening of an energy gap near the Fermi level at the  transition from the AFM$\perp$ state to the AFM$\parallel$
can be clearly seen in the corresponding transmission function $T(E)$ presented in Fig.~\ref{fig3}.

Numerical results for the whole range of the angle $\alpha$ (not presented), i.e. including also the hypothetical unstable magnetic states,
show that for a fixed value of $\lambda_{\rm so}$ the width of energy gap
strongly depends on the angle $\alpha$ as well as on the Hubbard parameter $U$.
Generally, the gap increases with $U$ and $\alpha$, independently of the nanoribbon width.
Interestingly, the system is non-magnetic for $\alpha=0$ and $U<1$,
but for higher values of $\alpha$ it exhibits the antiferromagnetic arrangement with quite considerable edge moments.
Transition to the AFM state opens an energy gap, and the gap width monotonically increases with $\alpha$.

\begin{figure}
\null\hskip -0.06\columnwidth
\includegraphics[width=0.97\columnwidth]{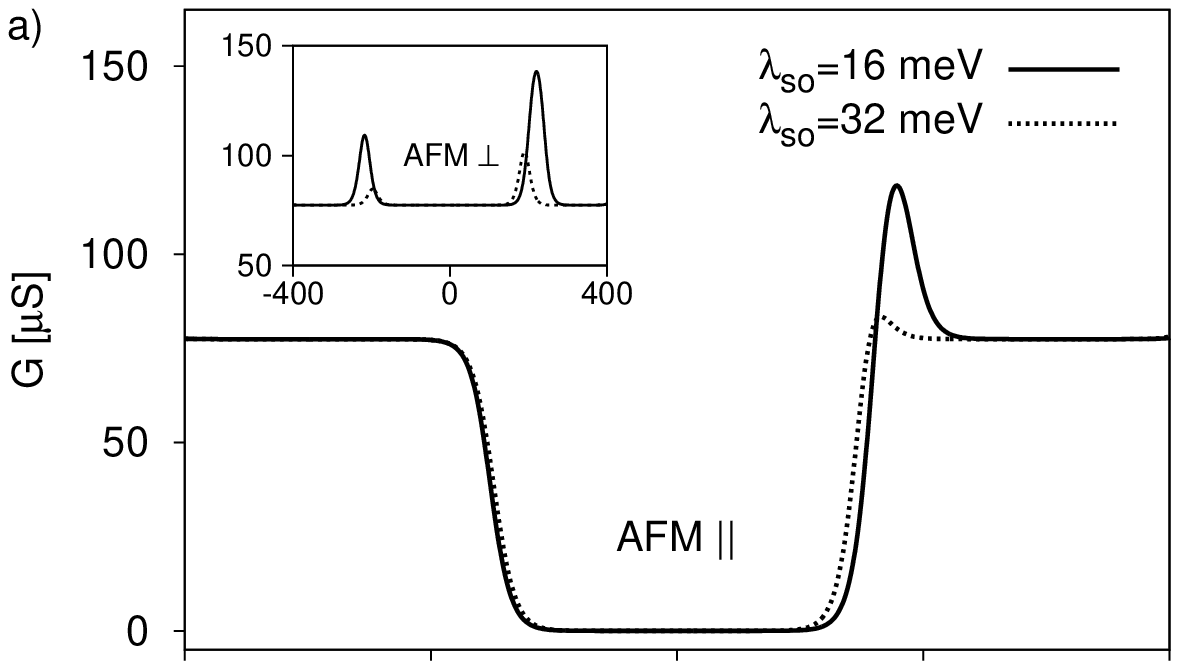}
\includegraphics[width=0.98\columnwidth]{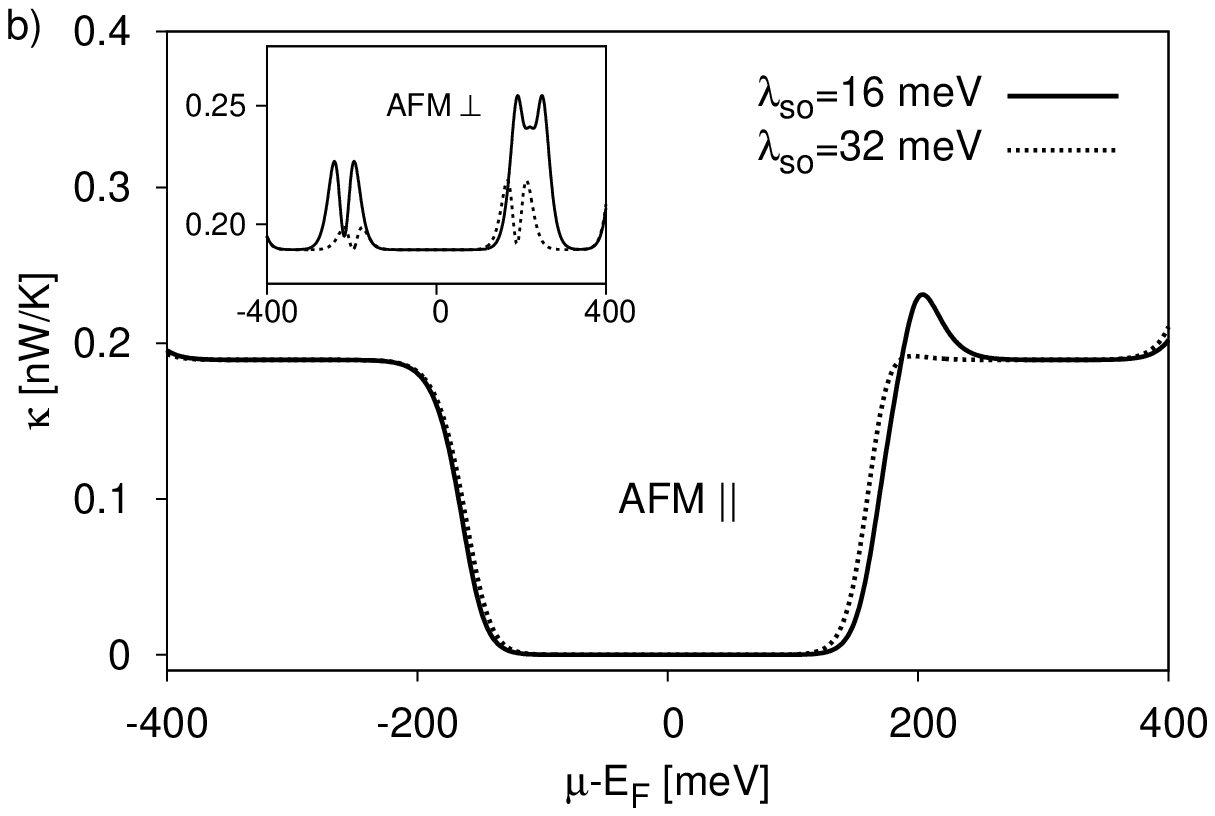}
\caption{\label{fig4}Electrical conductance $G$ (a) and electronic contribution to the thermal conductance, $\kappa$, (b) in the AFM$\parallel$ configuration, calculated as a function of the chemical potential for indicated values of  $\lambda_{\rm so}$ and $T=100\ \mbox{K}$. The other parameters as in Fig.2. The insets show the corresponding results for the AFM$\perp$ state.}
\end{figure}

Thermoelectric properties in the AFM$\perp$ phase were already analyzed in our earlier paper for a small value of $\lambda_{\rm so}$.\cite{Wierzbicki2015} Therefore, we focus here mainly on the AFM$\parallel$ ground state, though some new features in the thermelectric properties in the AFM$\perp$ state appear for large spin-orbit coupling, especially those associated with the transition from the conventional to topological insulator state.   The energy gap in the  AFM$\parallel$ phase has a significant influence on the transport and thermoelectric properties of the system.
The electrical conductance $G$ and the thermal conductance $\kappa$ due to electrons are practically equal to zero
in a wide energy region (corresponding to the gap) near the Fermi level $E_F$,
and then they strongly increase at the gap edges, see Fig.~\ref{fig4}. This figure shows both $G$ and $\kappa$ as a function of the chemical potential, and for both AFM$\parallel$ and  AFM$\perp$ (insets) phases. Interestingly, both $G$ and $\kappa$ in the AFM$\parallel$ state only weakly depend on $\lambda_{\rm so}$. This is rather obvious as the corresponding  gap in the spectrum (see  Figs.~\ref{fig2} and \ref{fig3}) only weakly depends on the spin-orbit coupling parameter.
Contrary, owing to the transition to the topological insulator state in the AFM$\perp$ phase, and closure of the gap for large enough spin-orbit coupling, both $G$ and $\kappa$ are constant and nonzero for chemical potentials in the region around $E_F$ (see the insets in Fig.~\ref{fig4}.
It is also worth to note, that in some narrow regions of the chemical potential, the electrical and thermal conductances in the AFM$\parallel$ and AFM$\perp$ phases are smaller for higher values of $\lambda_{\rm so}$.

\begin{figure}
\null\hskip -0.01\columnwidth
\includegraphics[width=0.99\columnwidth]{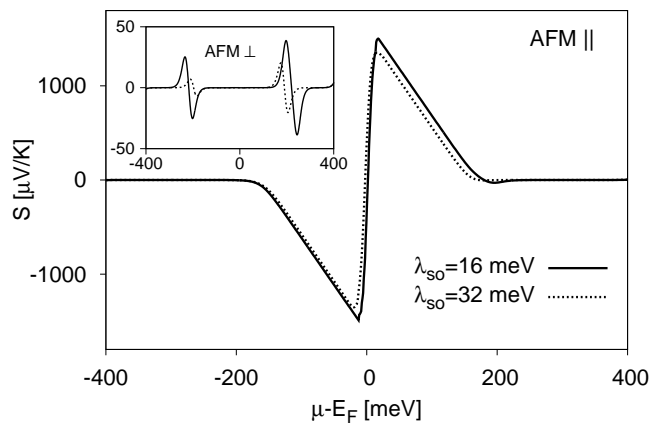}
\caption{\label{fig5} Thermopower $S$ for the AFM$\parallel$ configuration, calculated as a function of the chemical potential for  indicated values of $\lambda_{\rm so}$ and $T=100\ \mbox{K}$. The other parameters as in Fig.2. The inset shows the corresponding results for the AFM$\perp$ state.}
\end{figure}
\begin{figure}
\includegraphics[width=0.99\columnwidth]{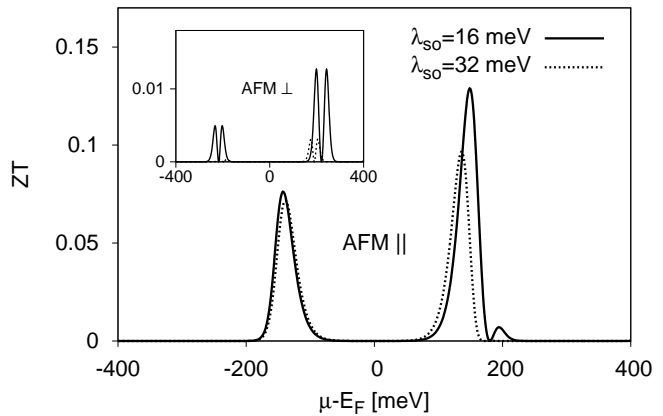}
\caption{\label{fig6} Figures of merit $ZT$ for the AFM$\parallel$ configuration, calculated as a function of the chemical potential for  indicated values of $\lambda_{\rm so}$  and $T=100\ \mbox{K}$. The other parameters as in Fig.2. The inset shows the corresponding results for the AFM$\perp$ state. }
\end{figure}

Due to the difference in the edge states, both  AFM$\perp$  and  AFM$\parallel$ phases display also  remarkably
different thermoelectric properties, especially the Seebeck coefficient (thermopower), see Fig.~\ref{fig5}. Due to the  gap in  the AFM$\parallel$ phase, see Fig.~\ref{fig3},
the thermopower $S$ considerably increases for chemical potentials in the vicinity of the Fermi level $F_F$, achieving very high values at low temperatures. Since the gap weakly depends on the spin-orbit coupling, the thermopower also weakly depends on  $\lambda_{\rm so}$. The situation in the AFM$\perp$ state is different. Now, the gap disappears for large values of $\lambda_{\rm so}$, so the thermopower remains zero in a broad region of chemical potentials around $E_F$, see the inset in  Fig.~\ref{fig5}. Then, smaller peaks appear due to the narrow peaks in transmission around $E\approx \pm 200$ meV, as shown in Fig.~\ref{fig3}.

The presented results show that nanoribbons with relatively strong spin-orbit coupling and exhibiting AFM$\parallel$ magnetic configuration
may be attractive for applications  for conversion of thermal into electrical energy.
The corresponding thermoelectric efficiency, described by the figure of merit $ZT$, is shown in Fig.~\ref{fig6}. Unfortunately, $ZT<1$, so the thermoelectric efficiency is below that anticipated for practical applications. However, one may conceive some ways of reducing the phonon heat conductance in order to increase $ZT$.

\subsection{Ferromagnetic phase}

Now, we  analyze ferromagnetic arrangement of the edge moments and compare the results with those obtained for the corresponding antiferromagnetic phase, presented and discussed in the  preceding section.
Both FM$\perp$ and FM$\parallel$ configurations are quasi-stable states with the corresponding energies above the energy of the AFM$\parallel$ ground state. The ferromagnetic configurations, however, can be stabilized externally by a magnetic field or due to proximity effects.

\begin{figure}
\includegraphics[width=0.54\columnwidth]{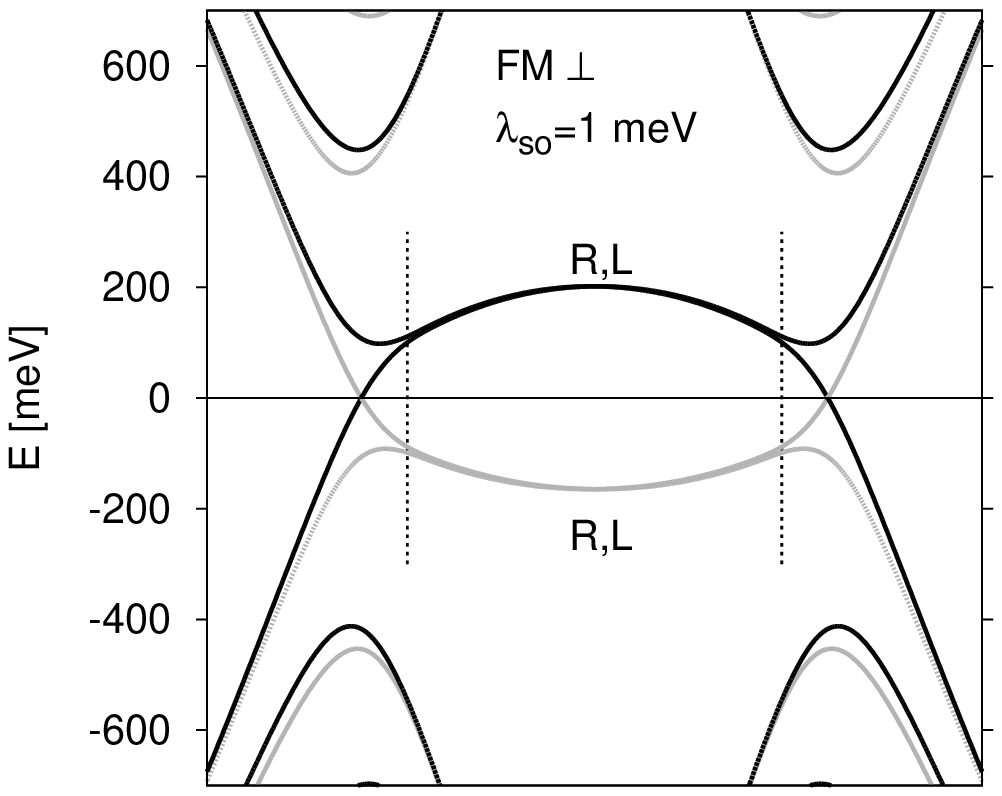}
\includegraphics[width=0.425\columnwidth]{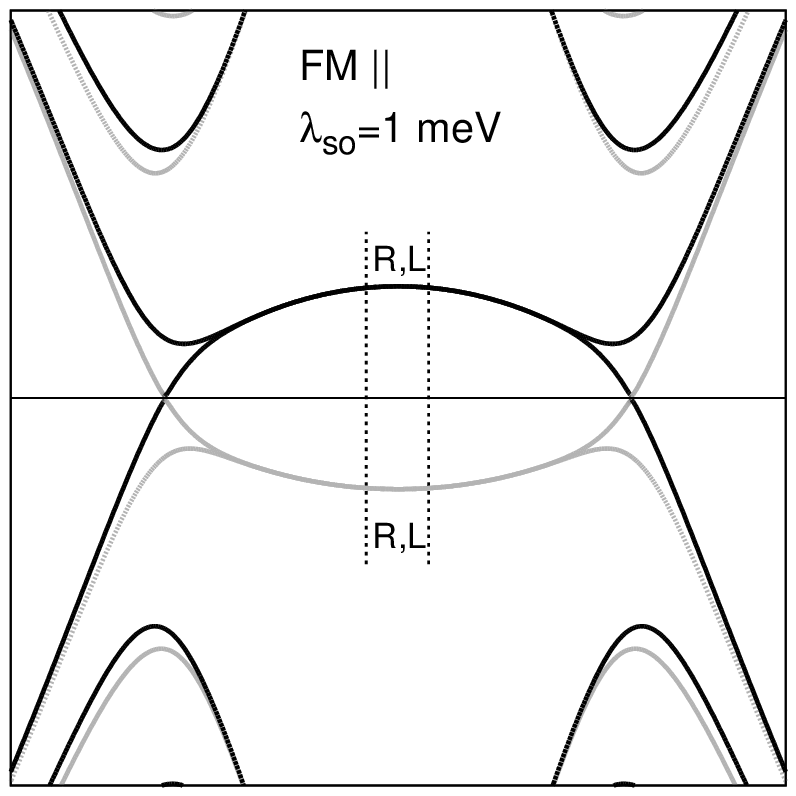}
\includegraphics[width=0.54\columnwidth]{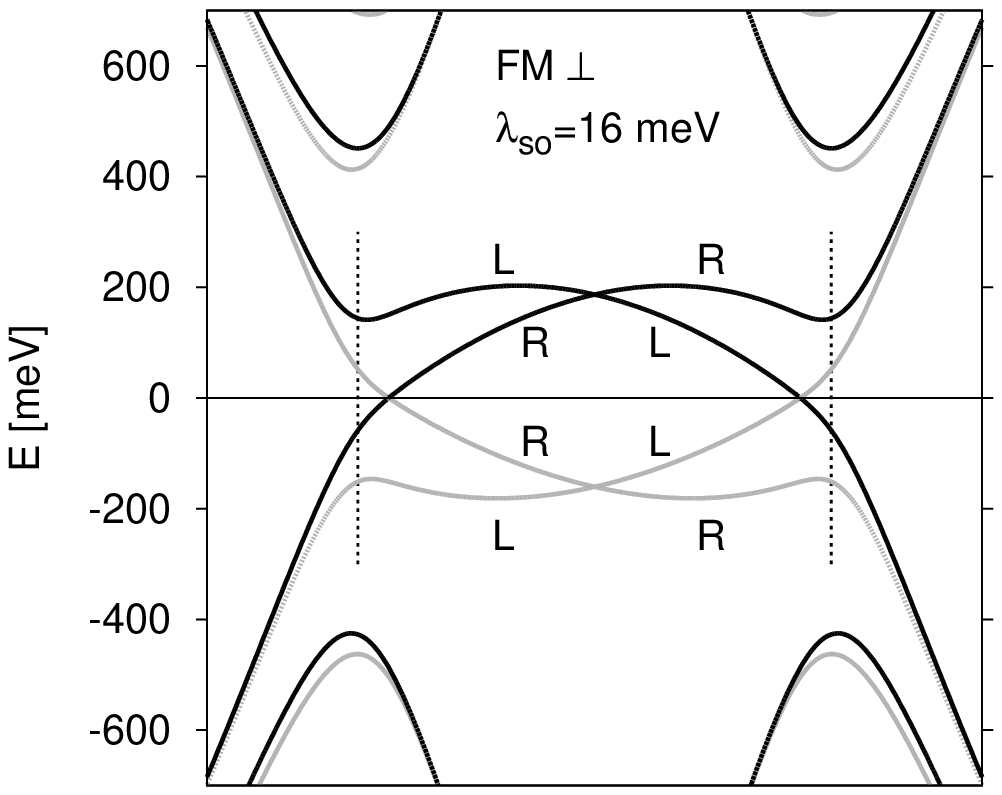}
\includegraphics[width=0.425\columnwidth]{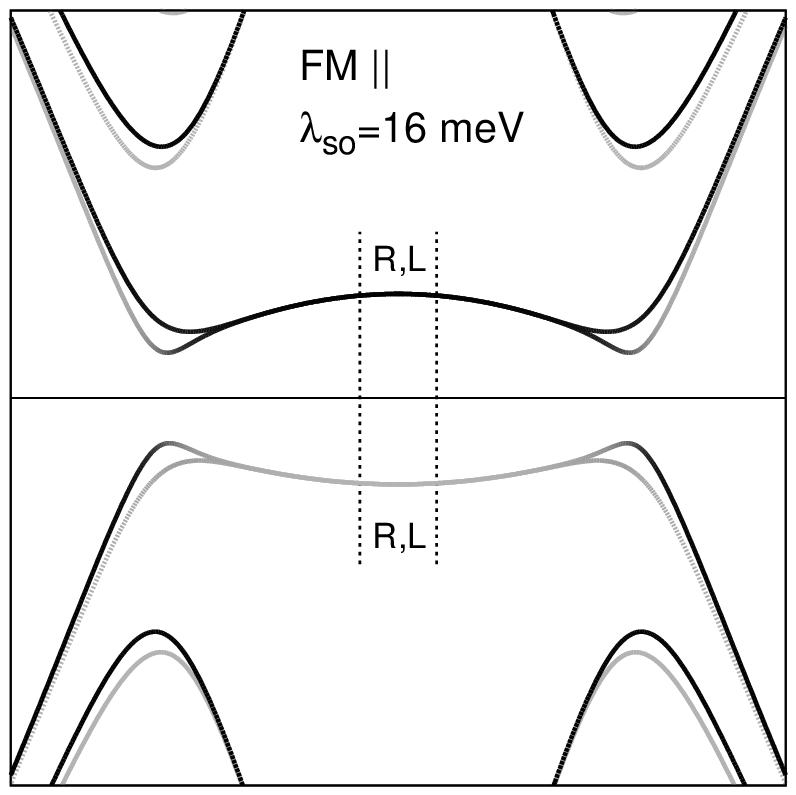}
\includegraphics[width=0.54\columnwidth]{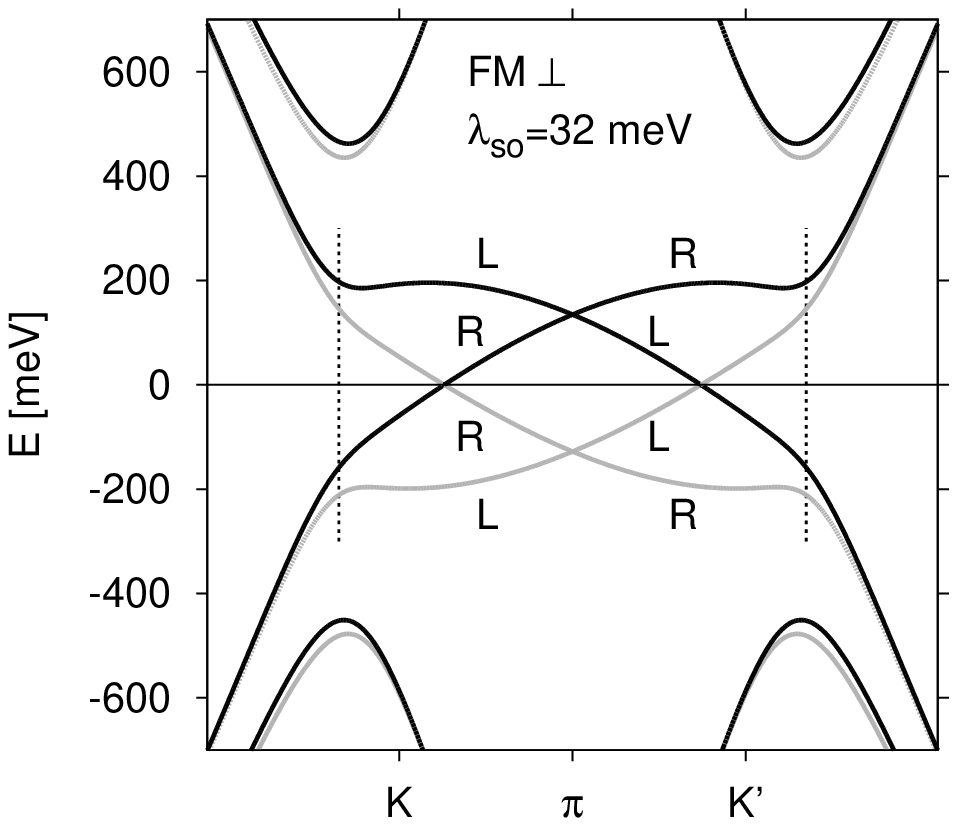}
\includegraphics[width=0.42\columnwidth]{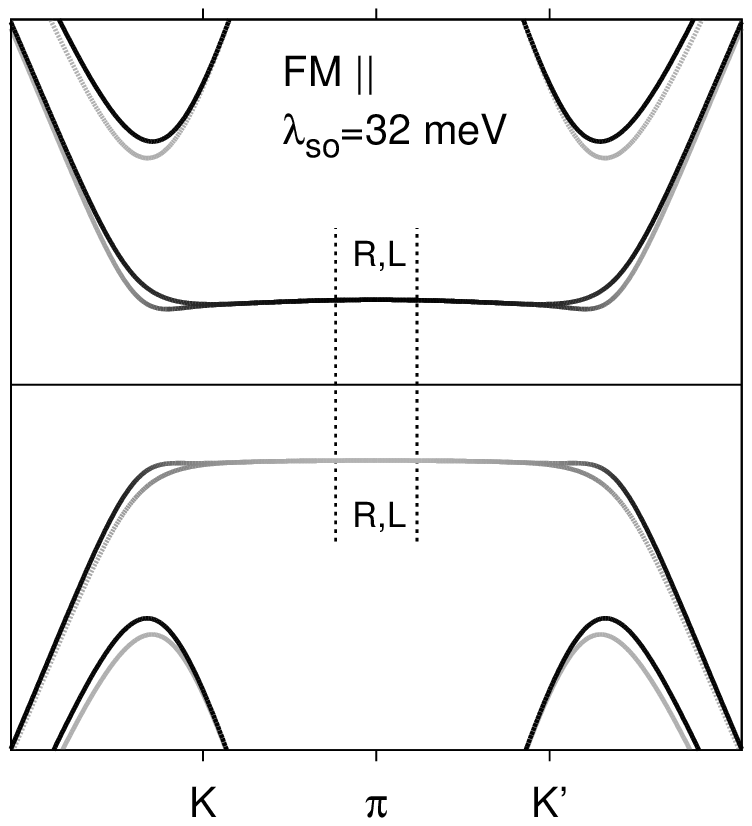}
\caption{\label{fig7} Band structure for the FM$\perp$ (left panel) and FM$\parallel$ (right panel) configurations.
Vertical lines denote the interval of wavevector $k$, where the edge parameter $\xi$ obeys the condition $\vert \xi \vert >0.5$. The assumed parameters are: $U=1.4$~eV, $N=8$, and $t=1.6$~eV.}
\end{figure}

\begin{figure}
\includegraphics[width=0.99\columnwidth]{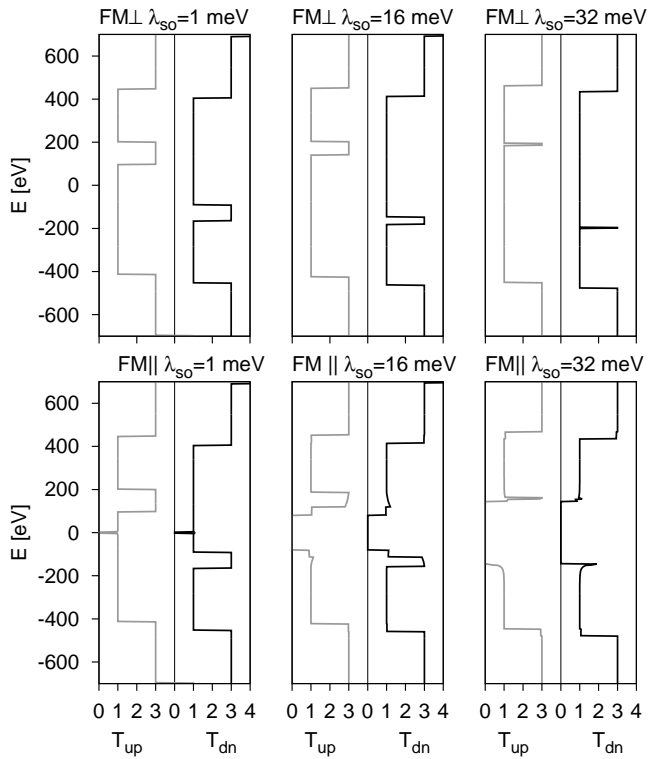}
\caption{\label{fig8} Transmission in the FM$\perp$ (top panel) and FM$\parallel$ (bottom panel) configurations, calculated for the same parameters as in Fig. 7.}
\end{figure}

The calculated band structure near the Fermi level $E_F$ is presented in Fig.~\ref{fig7}
for both FM$\perp$ and FM$\parallel$ phases,
whereas the corresponding spin-dependent transmission functions are shown in Fig.~{\ref{fig8}}.
As in the AFM configurations, electronic spectra of both FM$\perp$ and FM$\parallel$ phases are similar for small values of $\lambda_{\rm so}$, and both phases are in the topological insulator state. The situation changes for larger values  of $\lambda_{\rm so}$. According to Fig.~\ref{fig7}, the FM$\perp$ phase remains in the topological insulator state, independently of the spin-orbit coupling strength assumed in this figure.
However, for higher values of $\lambda_{\rm so}$ the degeneracy of states
localized at right and left edges of the nanoribbon is lifted.
Significantly  different qualitative behavior is obtained for the FM$\parallel$  phase.
Though the system is in the topological insulator phase for very small values of $\lambda_{\rm so}$, its character changes for stronger spin-orbit interaction,
and the system transforms into a conventional insulator with a relatively wide energy gap.
It is interesting to note, that the range of $k$-wavevectors, where
the edge states appear in the FM$\parallel$ phase, is much narrower than in the corresponding FM$\perp$ state
as well as in the previously discussed AFM phases
(see the dotted vertical lines in Fig.~\ref{fig7}).
Moreover, the states localized at the left and right edges are roughly degenerate.

The opening of an energy gap in the FM$\parallel$ phase with increasing $\lambda_{\rm so}$ and absence of such a gap in the FM$\perp$ state are well seen in Fig.~\ref{fig8},
where the spin-dependent transmission function is presented as a function of energy.
It is worth to note that the transmission function corresponding to a particular spin orientation, up or down,
is asymmetrical near the Fermi level,
i.e. transmission in the spin-up channel exhibits a peak  above $E_F$,
while the appropriate peak in the spin-down channel appears below the Fermi energy. This practically applies to both
FM$\parallel$ and FM$\perp$ configurations.
The width of the energy gap in the FM$\parallel$ state increases with increasing $\lambda_{\rm so}$.
However, no gap appears in the FM$\perp$ phase. Thus, the situation is remarkably different from that in the AFM configurations discussed above.

\begin{figure}
\includegraphics[width=0.99\columnwidth]{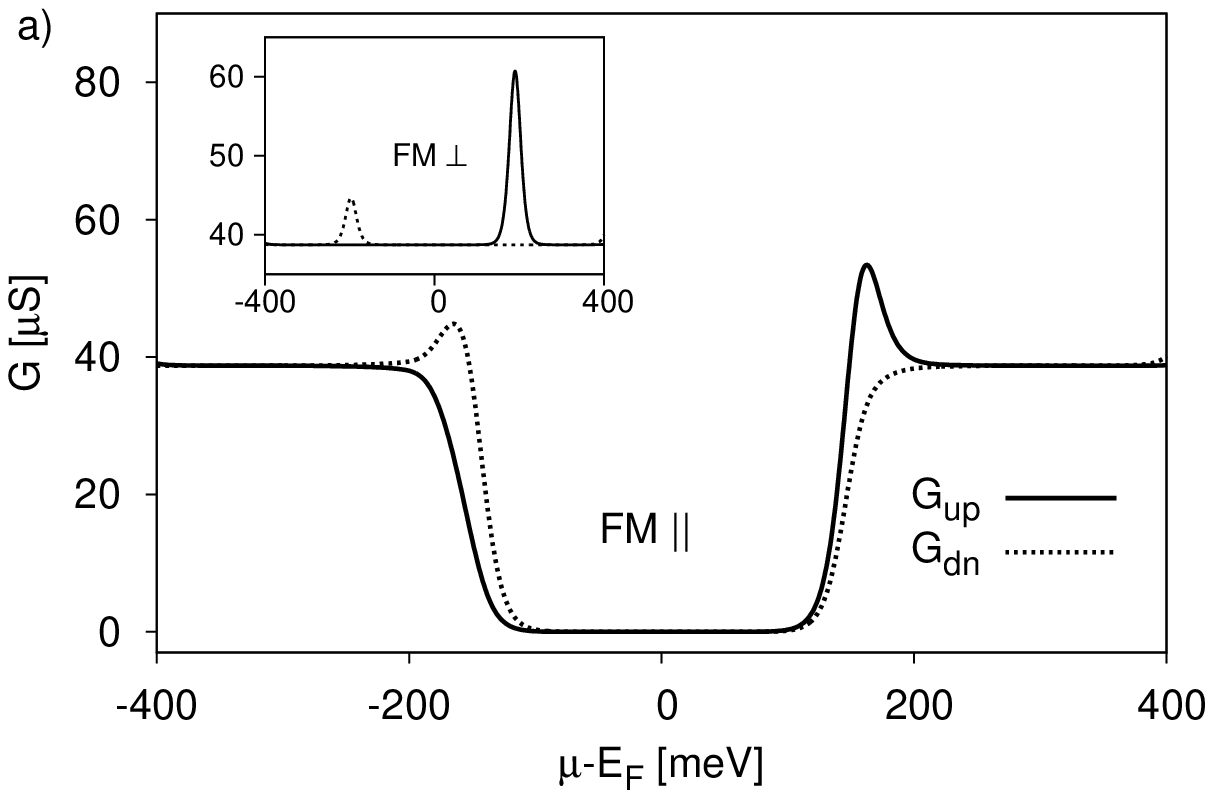}
\includegraphics[width=0.99\columnwidth]{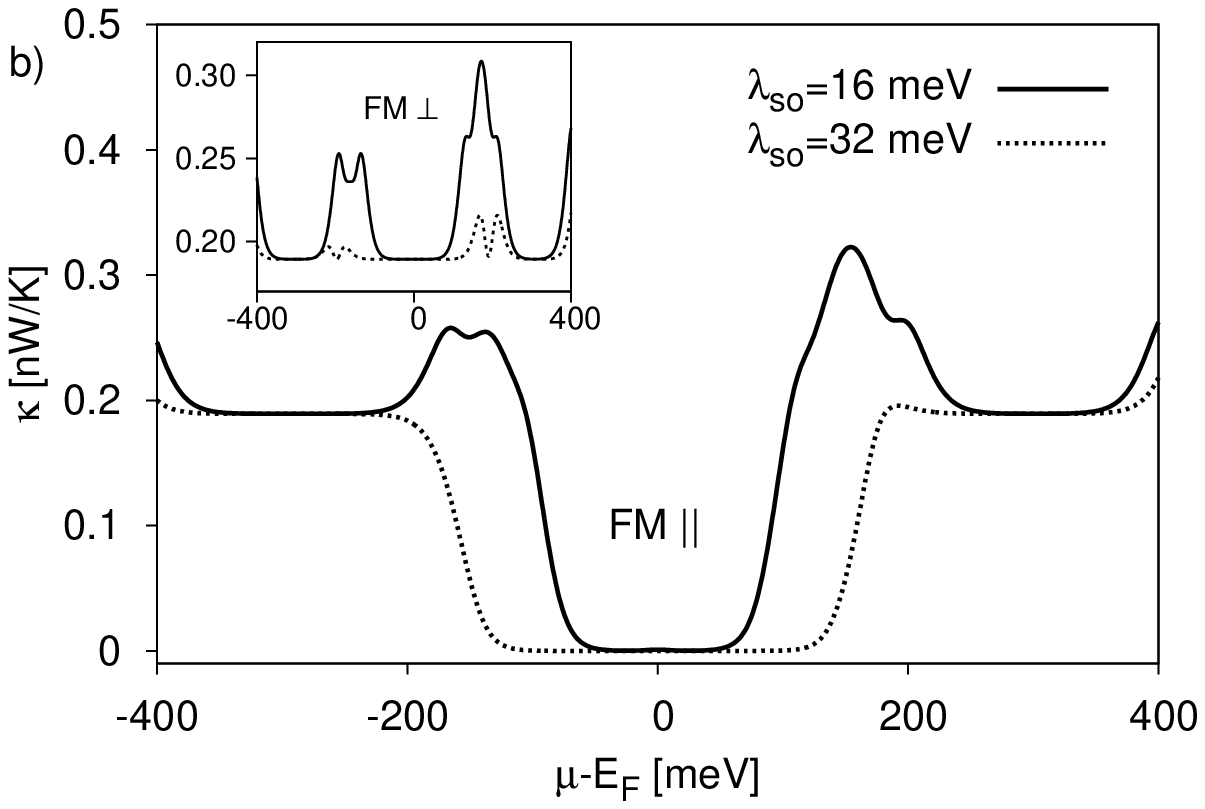}
\caption{\label{fig9} (a) Spin resolved electrical conductance $G$ in the FM$\parallel$ configuration for $\lambda_{\rm so}=32\ \mbox{meV}$. (b) Electronic contribution to the heat conductance $\kappa$ in the  FM$\parallel$ configuration for indicated values of $\lambda_{\rm so}$. The insets show the corresponding conductances in the FM$\perp$ state.
The results are for
$T=100\ \mbox{K}$ and other parameters as in Fig.7.}
\end{figure}
In the ferromagnetic phases, the two spin channels  contribute differently  to electronic transport.
If the spin relaxation time is long and therefore spin mixing is negligible, the spin-up and spin-down channels
can be treated as independent. One can then distinguish contributions of individual spin channels  to the electrical conductance (and also to the Seebeck coefficient).
The spin resolved electrical conductance is presented in Fig.~\ref{fig9}a as a function of chemical potential.
The wide central region with the conductance close to zero can be clearly seen for the FM$\parallel$ phase. This region
corresponds to the energy gap.
The conductance strongly increases near the gap edges, where it also remarkably depends on the spin orientation.
In turn, the conductance in the FM$\perp$ phase is finite and constant in the whole region of chemical potential shown in Fig.~\ref{fig9}a (see the  inset),
except the two peaks corresponding to the enhanced transmission in narrow regions of chemical potential for  the spin-up ($\mu>E_F$) and spin-down ($\mu<E_F$) channels, which appear near $E\approx\pm 200$ meV.

\begin{figure}
\null\hskip -0.03\columnwidth
\includegraphics[width=0.955\columnwidth]{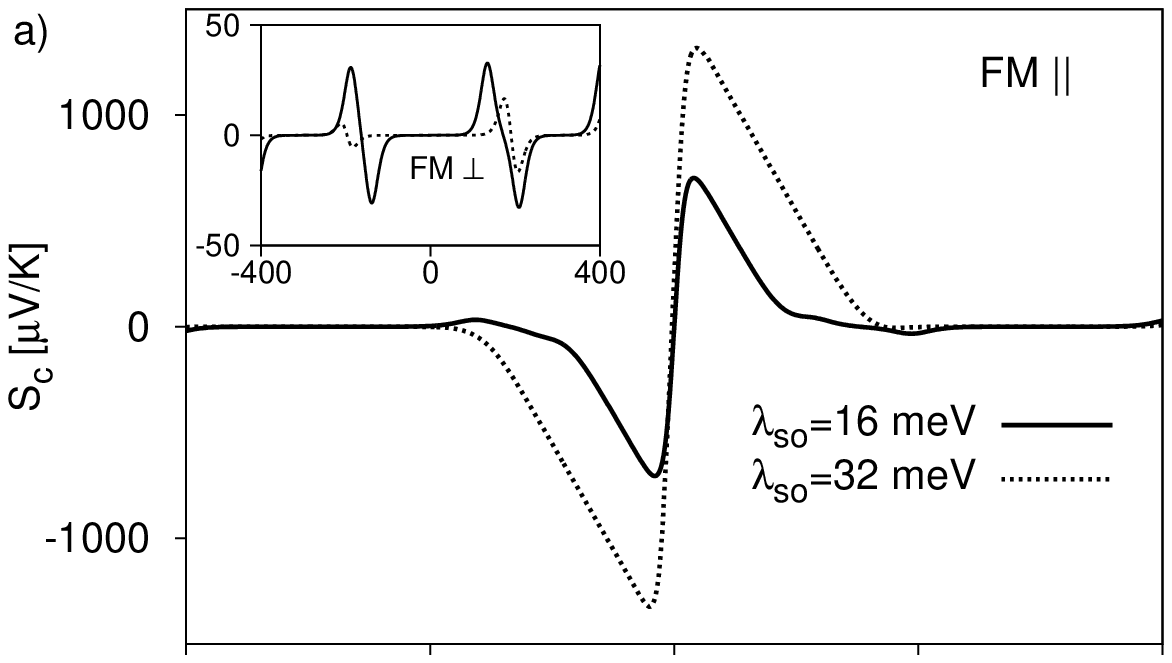}
\includegraphics[width=0.99\columnwidth]{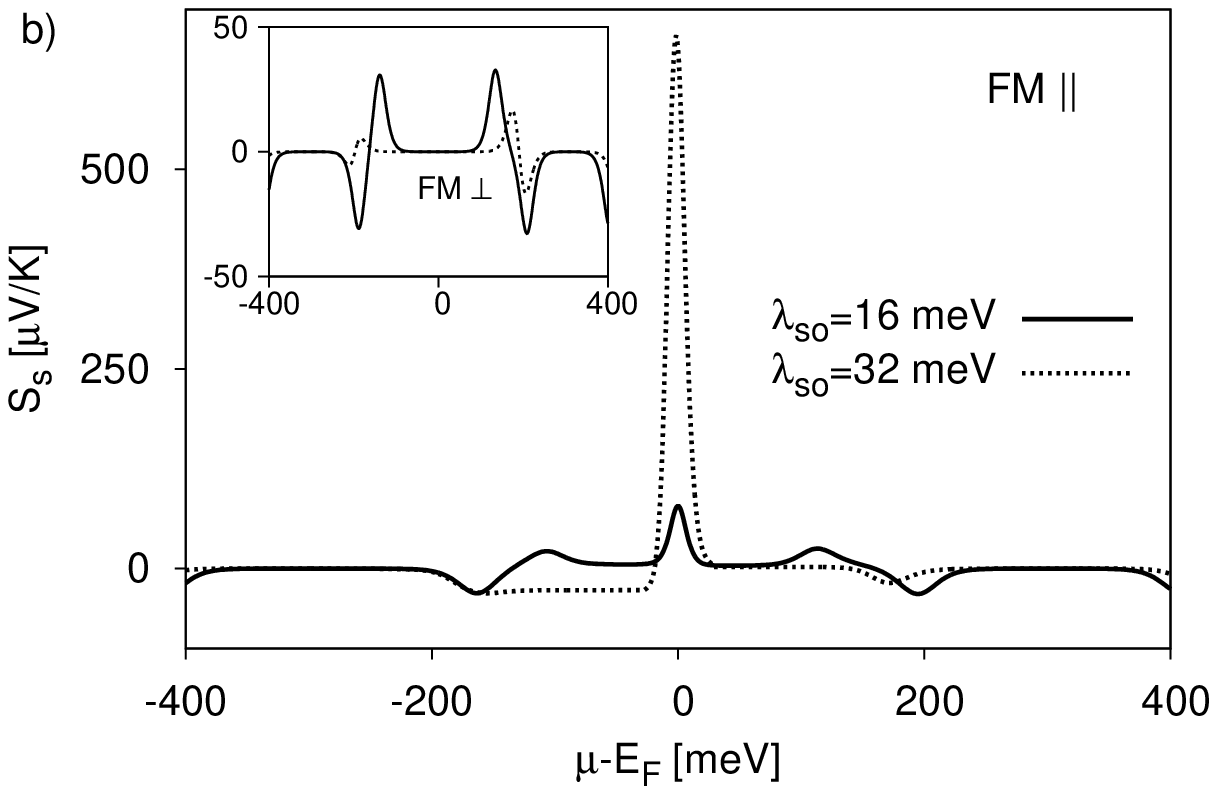}
\caption{\label{fig10} Conventional thermopower $S_c$ (a) and spin thermopower $S_s$ (b) for the  FM$\parallel$ configuration and indicated values of $\lambda_{\rm so}$. The insets show the corresponding thermopowers in the FM$\perp$ state. The results are for $T=100\ \mbox{K}$, and other parameters as in Fig.7}
\end{figure}
The thermal conductance due to electrons is shown in Fig.~\ref{fig9}b for two values of the spin-orbit parameter.
Interestingly, in some regions of chemical potential the thermal conductance is smaller for strong spin-orbit coupling,  see Fig.~\ref{fig9}b.
This appears mainly because the  energy gap in the FM$\parallel$ state is wider for stronger spin-orbit coupling, which effectively suppresses
$\kappa$ in a broader region of the chemical potential.
This difference is especially pronounced in the FM$\perp$ state (see the inset in Fig.~\ref{fig9}b),
where both peaks, i.e. the one  below and another one above the Fermi level, are notably reduced for the larger value of the spin-obit parameter $\lambda_{\rm so}$.

\begin{figure}
\null\hskip 0.03\columnwidth
\includegraphics[width=0.95\columnwidth]{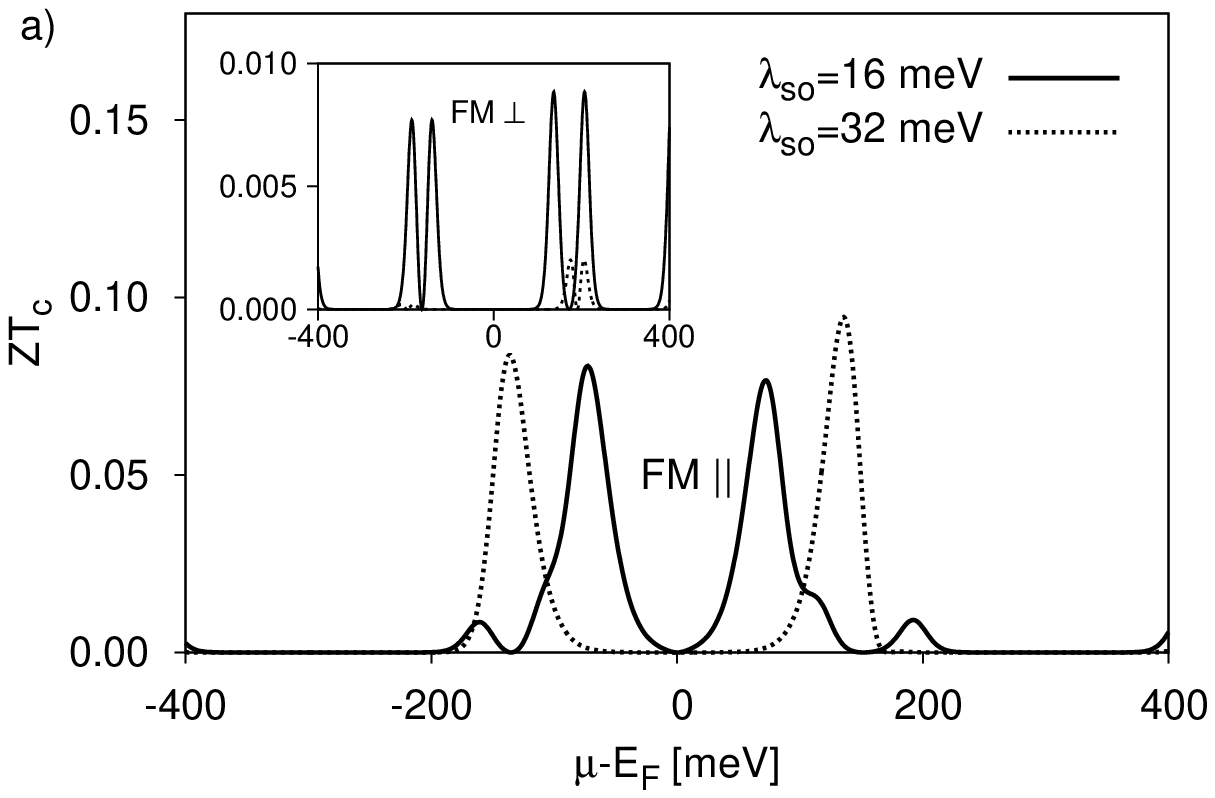}
\includegraphics[width=0.99\columnwidth]{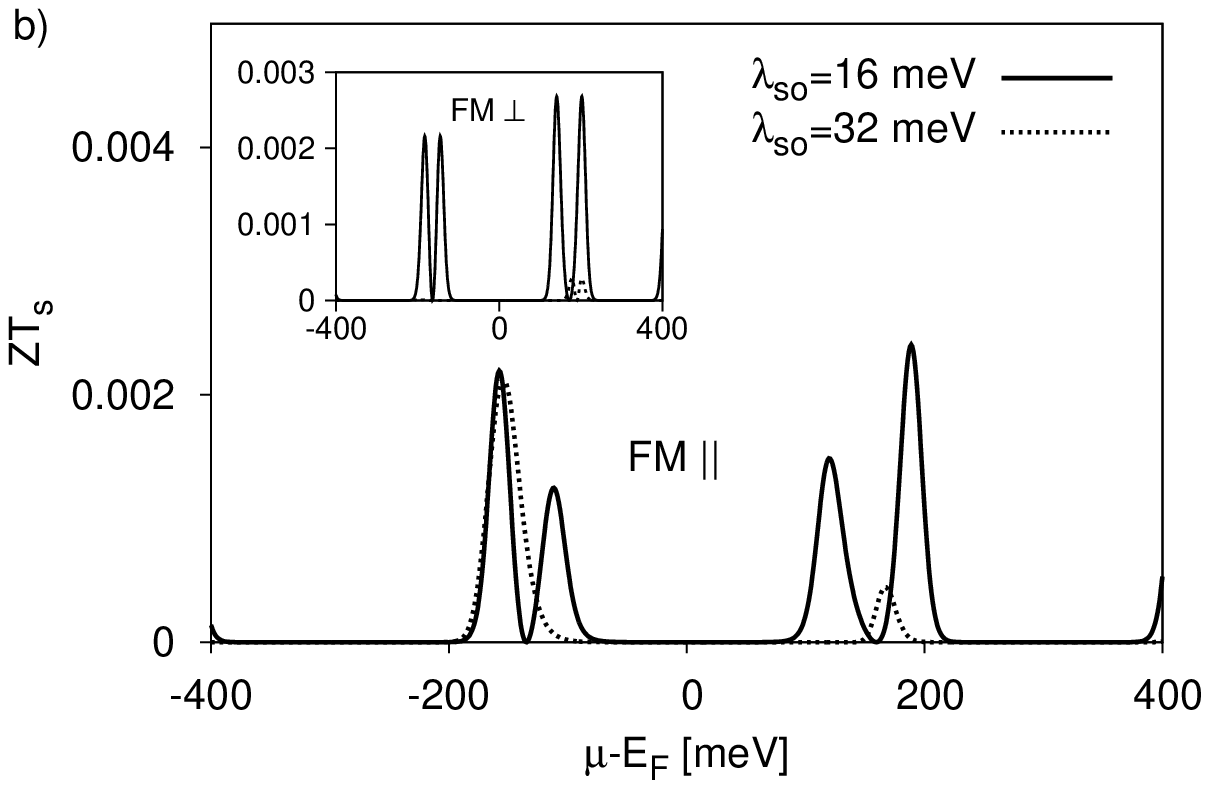}
\caption{\label{fig11} Figure of merit for the conventional thermoelectricity,  $ZT_c$, (a) and for the spin thermoelectricity, $ZT_s$, (b) in the  FM$\parallel$ configuration and for indicated values of the spin-orbit coupling. The insets show the corresponding figures of merit  for the FM$\perp$ state. The results are for $T=100\ \mbox{K}$, and other parameters as in Fig.7.}
\end{figure}
In addition to the conventional thermopower, one can now also define the  spin thermopower. Both the charge $S_c$ and spin $S_s$ Seebeck coefficients
are presented in Fig.~\ref{fig10} for two values of the parameter $\lambda_{\rm so}$.
Due to the energy gap in the FM$\parallel$ state, the conventional Seebeck coefficient $S_c$
is considerably enhanced in the vicinity of the Fermi level.
The enhancement strongly depends on $\lambda_{\rm so}$ and increases for higher values of the spin-orbit coupling.
Similarly, the spin thermopower $S_s$ also exhibits a very pronounced peak for a narrow region of chemical potential near $E_F$,
and then remains rather small for chemical potentials outside this region.
The peak in $S_s$ is very narrow, much narrower than the relevant energy gap.
Furthermore, height of the peak strongly depends on the spin-orbit parameter.
Remarkably different behavior has been obtained for the FM$\perp$ phase which exhibits typical topological insulator features.
Now, the $S_c$ and $S_s$ Seebeck coefficients are rather independent of the chemical potential in the vicinity of the Fermi level, as shown in the insets to Fig.~\ref{fig10}). The small peaks
which appear for higher values of $\vert \mu - E_F\vert$  follow from the narrow peaks in the corresponding transmission functions, see Fig.~\ref{fig8}.

The corresponding figures of merit for the conventional and spin thermoelectricity are shown in
Figs.~\ref{fig11}a and \ref{fig11}b, respectively. Significant values of the figures of merit occur only in certain regions of the chemical potential, where the corresponding thermopowers reach local maxima. Both of them, however, are relatively small, similarly as it was in the case of AFM configurations. However, especially small is the spin figure of merit, even if the corresponding spin thermopower is relatively high. This is because the spin thermoelectric efficiency in the narrow region of chemical potential around $E_F$,  where the relatively high peak appears in the spin thermopower, is negligible (compare Fig.~\ref{fig10}b and Fig.~\ref{fig11}b). This suppression of $ZT_s$ appears  due to vanishing spin conductance in this region of chemical potential (note, the transmission functions for both spin orientations are equal this region, see Fig.~\ref{fig8}).

\section{Summary and conclusions}

We have analyzed the influence of Coulomb and spin-orbit interactions in  nanoribbons of 2D buckled hexagonal crystals with zigzag edges on the edge magnetic moments and on the
topological  and thermoelectric properties. We have shown that the Coulomb interaction, taken in the form of Hubbard term in the mean field approximation, leads to formation of the edge magnetic moments, which for a sufficiently strong spin-orbit coupling  are oriented  in the nanoribbon plane  in the ground state, and the moments at one edge are opposite to those at the other edge (AFM$\parallel$ phase) -- in agreement with Lado et al.\cite{Lado2014} For small spin-orbit coupling, this configuration is almost degenerate with the state, in which magnetic moments are oriented perpendicularly to the nanoribbon plane (AFM$\perp$ phase). States with ferromagnetic arrangements of the moments (FM$\parallel$ and FM$\perp$ phases) are quasi-stable and correspond to  higher energies. Such states can be stabilized externally, e.g. by a magnetic field and/or due to proximity to a ferromagnetic contact.

As the AFM$\parallel$ phase remains in the topological insulator state when the pin-orbit coupling increases, the AFM$\perp$ phase undergoes then a transition from the topological to conventional insulator, i.e. an energy gap opens in the corresponding edge states. This gap has a significant influence on transport and thermoelectric properties. First of all, the opening of the gap is associated with a nonzero thermopower in the vicinity of the gap edges. The ferromagnetic quasi-stable states, FM$\parallel$ and FM$\perp$, behave differently with increasing $\lambda_{\rm so}$, as well. This behavior qualitatively resembles that of the antiferromagnetic phases, but now the gap in the FM$\parallel$ state, which opens with increasing $\lambda_{\rm so}$, is spin dependent. As a result, a nonzero spin thermopower may be observed, in addition to the conventional one.

\begin{acknowledgments}
This work was supported by the National Science Center in Poland as Project No.~DEC-2012/04/A/ST3/00372.
\end{acknowledgments}

\end{document}